# A Bayesian Hierarchical Framework for Capturing Preference Heterogeneity in Migration Flows

Aric Cutuli[1], Upmanu Lall[2,3], Michael J. Puma[1,4], Émile Esmaili[1], and Rachata Muneepeerakul[5]

[1]Center for Climate Systems Research, Columbia Climate School, Columbia University, New York, NY, USA
[2]Columbia Water Center, Columbia University, New York, NY, USA
[3]Water Institute, School of Complex Adaptive Systems, Arizona State University, Tempe, AZ, USA
[4]NASA Goddard Institute for Space Studies, Columbia Climate School, Columbia University, New York, NY 10025
[5]Department of Agricultural and Biological Engineering, University of Florida, Gainesville, FL, United States of America

## ABSTRACT

Understanding and predicting human migration patterns is a central challenge in population dynamics research. Traditional physics-inspired gravity and radiation models represent migration flows as functions of attractiveness using socio-economic features as proxies. They assume that the relationship between features and migration is spatially invariant, regardless of the origin and destination locations of migrants. We use Bayesian hierarchical models to demonstrate that migrant preferences likely vary based on geographical context, specifically the origin-destination pair. By applying these models to U.S. interstate migration data, we show that incorporating heterogeneity in a single latent migration parameter significantly improves the ability to explain variations in migrant flows. Accounting for such heterogeneity enables it to outperform classical methods and recent machine-learning approaches. A clustering analysis of spatially varying parameters reveals two distinct groups of migration paths. Individuals migrating along low-flow paths (typically between smaller populations or over larger distances) exhibit more nuanced decision-making. Their choices are less directly influenced by specific destination characteristics such as housing costs, land area, and climate-related disaster costs. High-flow path migrants appear to respond more directly to these destination attributes. Our results challenge assumptions of uniform preferences and underscore the value of capturing heterogeneity in migration models and policymaking.

## INTRODUCTION

Models that can reliably forecast human migration flows have become indispensable tools across many domains, including city and infrastructure planning, international trade, conservation planning, public policy, and mitigating the spread of infectious diseases (Bengtsson et al., 2015; Marshall et al., 2018; Tizzoni et al., 2014). Similarly, accurate commuting models are vital tools for urban planning and policy making. They inform infrastructure investment, transportation policies, and zoning decisions to mitigate issues like traffic congestion and pollution as communities evolve (De Montis, Chessa, Campagna, Caschili, & Deplano, 2010; Zhang, Xu, Tu, & Ratti, 2018).

While migration and commuting models have many important applications, a key challenge lies in the data used to build these models (Rampazzo, Rango, & Weber, 2023). Flow statistics are typically reported with a high degree of epistemic uncertainty, as data is often generated through auxiliary procedures using stock data or by surveying population samples (Abel & Cohen, 2019; Ghimire, Williams, Thornton, Young-DeMarco, & Bhandari, 2019). Moreover, it is documented that demographic features and perhaps unobserved regional characteristics can impact empirical relationships between covariates and the propensity for individuals to migrate, suggesting the presence of heterogeneities in flow data (Cattaneo & Peri, 2016). Despite this evidence of heterogeneities in migration and mobility patterns, the majority of studies on flow forecasting restrict model parameters to be common across region pairs, disregarding potentially systematic spatial variations in the underlying migration dynamics across regions.

To address these shortcomings, we propose a Bayesian hierarchical framework that permits the regression parameters to vary spatially, while effectively managing the variability in regression parameters through spatial pooling techniques. For a system with $N$ regions in which we consider bilateral migration, this entails replacing a regression parameter $\theta$ by a set of $N(N-1)$ parameters $\{\theta_{i,j}\}$, which gives the number of unique pairs (i.e., bilateral relationships) in a system with $N$ regions. This set of parameters is used to model flow data generated by each origin-destination pair $(i,j)$, accounting for the distinct migration dynamics between each pair. In this paper, we introduce two types of models: one where the spatial variation is applied to the parameter governing latent migration (e.g., the intercept term in a linear regression), and another where spatial variation extends to all parameters. Recently, Welch and Raftery (2022) utilized this concept to model migration endogenously, recognizing that migration patterns can both influence and be influenced by other factors within the same geographical and social systems. In this paper, we make use of a hierarchical paradigm to enhance traditional migration models, which typically rely on a limited set of exogenous variables.

Our developed hierarchical models outperform the traditional methods in predicting migration flows, demonstrating their effectiveness in capturing complex migration dynamics. We cluster spatially varying parameters to elicit insights into the heterogeneity of migrants' preferred destination characteristics. Moreover, while confirming traditional theories that state pairs with low migration flow are generally smaller in population and further apart, our analysis reveals that individuals migrating between low-flow states place less emphasis on housing prices, state size, and climate-related characteristics compared to those migrating between high-flow states.

**Traditional Models**

The gravity model, derived from Newton's law of gravity and introduced in its modern form in Zipf (1946), considers the migration between two locations to be proportional to the product of their individual utilities and inversely proportional to the distance that separates them. In its simplest formulation, the utility of each location is defined by population.

$$M_{i,j,t} = \kappa \frac{P_{i,t} P_{j,t}}{D_{i,j}} \quad (1)$$

where $M_{i,j,t}$ is the migration flow from location $i$ to location $j$ at time $t$, $\kappa$ is a proportionality constant, $P_{i,t}$ is the population of location $i$ at time $t$, $P_{j,t}$ is the population of location $j$ at time $t$, and $D_{i,j}$ is the distance between locations $i$ and $j$.

The radiation model, introduced by Simini, Gonzalez, Maritan, and Barabási (2012), has been shown to give better empirical predictions of migration flows of commuters. The model frames individuals as particles traveling through space with some probability of being absorbed by competing venues. For an individual making a trip from zone $i$ to zone $j$, this probability is interpreted as directly proportional to the level of opportunity, or the aggregate utility, that intermediate zones offer. The average flux, or predicted flow, from $i$ to $j$ is

$$M_{i,j,t} = M_{i,t} \frac{P_{i,t} P_{j,t}}{(P_{i,t} + S_{i,j,t})(P_{i,t} + P_{j,t} + S_{i,j,t})}, \qquad (2)$$

where $M_{i,t}$ is the sum of individuals leaving zone $i$, and $S_{i,j,t}$ represents the total population of intervening regions, defined as $S_{i,j,t} = \sum_k P_{k,t}$ where $D_{i,k} < D_{i,j}$ For performance reasons, $M_{i,t}$ is assumed to be proportional to the population of zone $i$, such that $M_{i,t} = \kappa P_{i,t}$.

Rather than assuming unit exponents for the features, estimation of these models involves taking the logarithm of both sides and fitting

$$\log M_{i,j,t} = \alpha + \beta_1 \log P_{i,t} + \beta_2 \log P_{j,t} + \beta_3 \log D_{i,j} + \epsilon_t \quad (3)$$

and

$$\log M_{i,j,t} = \alpha + \beta_1 \log P_{i,t} + \beta_2 \log P_{j,t} + \beta_3 \log(P_{i,t} + S_{i,j,t}) + \beta_4 \log(P_{i,t} + P_{j,t} + S_{i,j,t}) + \epsilon_t \qquad (4)$$

respectively, via traditional statistical methods such as ordinary least squares or maximum likelihood estimation. Here, $\alpha = \log \kappa$ and $\epsilon_t$ is a white-noise process with a mean of 0. For brevity, we will occasionally refer to the collection of these model parameters as θ.

### ADOPTING A HIERARCHICAL BAYESIAN FRAMEWORK

The main advantages of the traditional models are their interpretability and ease of estimation. However, these models are limited in their ability to accurately reflect the complex relationships between features and migration flows that vary uniquely across different region pairs or contextual settings. Essentially, by estimating common rather than context-specific parameters (i.e., θ rather than θ$_{i,j}$), it is assumed that each observation emerges from an identical distribution. We consider that there may be systematic variation in the data-generating process, which can be captured by estimating context-specific parameters in a hierarchical Bayesian framework.

In a hierarchical model (Figure 1), the interactions among the hierarchical tiers facilitate mutual learning across contexts without compromising their distinct nuances (Betancourt, 2020). Parameters are allowed to vary spatially at one level, but their variation is "shrunk" to a common distribution or informed by measurable attributes at the next level, a scheme commonly referred to as partial pooling. Full pooling (equivalent to the treatment in traditional models) and no pooling (unconstrained spatial variation in parameters) are two extremes of this approach. Partial pooling strikes a balance between these extremes, where the appropriate degree of pooling is determined by the data itself, thus providing the opportunity to explore departures from the traditional model as well as situations where departures are most pronounced. In underpowered studies (e.g., migration modeling) where uncertainty is high, this is particularly helpful, as partial pooling allows for the shrinkage of context-specific effects toward common effects in contexts where uncertainty is high, which ameliorates issues of sign and magnitude error (Gelman, Hill, & Yajima, 2012). This can effectively highlight certain contexts with similar or highly dissimilar posterior coefficients for selected predictors.

[INSERT FIGURE 1 HERE]

In this paper, we compare two hierarchical parameterizations for each model as they are written in (3) and (4), which we will refer to as "hierarchical gravity" and "hierarchical radiation" models. One specification considers spatial variation over the intercept term, while the other considers spatial variation over all parameters (Figure 1). In the spatially varying specification, each coefficient becomes location-pair specific (denoted with i,j subscripts, e.g., $\beta_{1:i,j}$, $\beta_{2:i,j}$, $\beta_{3:i,j}$ representing the coefficients for origin population, destination population, and distance, respectively). However, in the estimation of the hierarchical gravity model with all parameters varying, we let $\beta_{3:i,j}$ (the coefficient for distance) be common across all $i, j$, because the corresponding covariate $D_{i,j}$ is not time-varying; allowing for spatial variation in $\beta_3$ would introduce collinearity with the varying intercept parameter. In our initial exploration of the data, we observe heteroskedasticity when estimating (3) and (4) via ordinary least squares, so we consider a prior for the variance parameters to account for heterogeneity through partial pooling.

Our specification employs informative location priors to efficiently pinpoint the central tendencies of universal effects and uses wider variance priors to allow for efficient exploration of the effect space across different contexts. In particular, common effect hyperparameters (e.g., $\hat{\mu}_{\beta_k}, \hat{\sigma}_{\beta_k}, \hat{\mu}_\alpha, \hat{\sigma}_\alpha, \hat{\mu}_\sigma$) are set as the ordinary least squares estimates of (3) and (4), since these estimates are analogs for common effects in the hierarchical setting. Prior scales of common effects are given hyperparameters large enough to encourage the exploration of context-specific parameter spaces. Since we pre-define log $M_{i,j,t} \geq 0$ for all $(i, j, t)$, we use a truncated normal distribution to define an appropriate support for the likelihood.

## MODELING INTERSTATE MIGRATION FLOW IN THE UNITED STATES

Building on the hierarchical Bayesian framework outlined previously, we apply these models to the task of analyzing interstate migration flows in the United States. This application allows us to compare our Bayesian hierarchical models to their classical counterparts as well as to recent machine learning techniques.

To conduct this analysis, we use publicly available data from US Census Bureau surveys, which estimate the number of individuals migrating between the 51 states (including the District of Columbia) from 2005 through 2019 (Census Bureau, 2024). The data is given in the form of 90% confidence intervals, which we use to sample model performance across paths of plausibly true migration counts. Each model is systematically trained using the same dataset, and performance metrics are derived from several test sets that represent plausible migration counts. Population numbers also come from US Census data (Census Bureau, 2021), and the distance between states is computed as the great-circle distance between the coordinates of capital cities.

Additional data sources that are used in a later clustering analysis include climate-related disaster costs (in billions of US dollars) from the National Centers for Environmental Information (NCEI, 2024), land area (in square miles), and a housing index (percentage) from the Federal Housing Finance

Agency (FHFA, 2024) that measures the percent deviation from 1990 aggregate housing price levels for each state.

**Experimental Setup**

We evaluate the performance of various models by predicting migration flows for the years 2017 through 2019 using data from 2005 through 2016. To avoid issues with multimodality, we exclude cases where migration flows are zero, focusing our analysis on state pairs with non-zero migration flows. We acknowledge that our sampling of observations underestimates the true variance in migrant counts, as we do not know the sample size of the survey; nonetheless, the procedure provides valuable insights into uncertainty for the different models.

*Model Estimation*

In accordance with traditional methods, we fit the simple gravity and radiation models are fit using ordinary least squares. Although we recognize that this approach does not fully satisfy classic assumptions such as homoskedasticity and uncorrelated residuals, we proceed nonetheless. The estimation of our Bayesian models is done using the No-U-Turn Sampling (NUTS) algorithm, which currently sits at the frontier of approximate Bayesian inference methods that leverage Hamiltonian flows (Hoffman & Gelman, 2014). We also make use of reparameterization techniques available to Google's NumPyro probabilistic programming library, such as de-centering context-specific Gaussian priors and mapping from variationally inferred parameterizations to ensure geometrically ergodic sampling spaces (Gorinova, Moore, & Hoffman, 2020; Phan, Pradhan, & Jankowiak, 2019).

*Machine Learning*

The linear design of the gravity and radiation models sacrifices predictive power for greater interpretability. Due to the complexity of migration dynamics, pinning down any one set of features to precisely explain them is an intractable goal. Hence, while the interpretability of such traditional model structures is an attractive property in a policy-making and robustness context, the flexibility and performance benefits of non-linear methods are a worthy alternative if we seek to solely maximize predictive capability. Robinson and Dilkina (2018) showcased the ability of "extreme" gradient boosting (XGBoost) and feedforward neural networks (ANNs) to outperform the gravity and radiation approaches for international and intercounty migration flow prediction (Chen & Guestrin, 2016; LeCun, Bengio, & Hinton, 2015). We introduce these models as competitors in our analysis.

In the models, we include all gravity and radiation features: $P_{i,t}$, $P_{j,t}$, $P_{i,t}+S_{i,j,t}$, $P_{i,t}+P_{j,t}+S_{i,j,t}$, and $D_{i,j}$. All features are logarithmically transformed and subsequently scaled using the logistic z-score method

$$x \to \frac{1}{1+\exp[-(x-\bar{x})/\sigma]}, \qquad (5)$$

which drives features to be within the unit interval. Hyperparameters are tuned through a 5-fold randomized search over 50 combinations in which we iteratively drop one year of data from the training set, leave the dropped year as a validation set, and select the best hyperparameters evaluated with root-mean-square error. For the XGBoost models, we consider a variety of maximum tree depths in the range [2,10], gradient-boosted tree quantities in [50,350], and learning rates in [0.01,0.5]. For the ANN, we consider hidden layer quantities in [1,8], widths in [16,128], and minibatch sizes in $\{2^4, 2^5, 2^6, 2^7, 2^8\}$, and use early stopping to moderate convergence.

*Evaluation Metrics*

To assess the performance of competing models, we employ four out-of-sample metrics that measure the accuracy of a predicted migration matrix in reproducing the true values. The first two metrics are mean absolute error (MAE) and $R^2$, while the latter two are similarity scores commonly utilized in previous studies for evaluating human mobility models (Lenormand, Bassolas, & Ramasco, 2016; Lenormand, Huet, Gargiulo, & Deffuant, 2012; Robinson & Dilkina, 2018). The Common Part of Commuters (CPC) is a commuting analog of the Bray-Curtis similarity score that compares the predicted and observed migration flow (Faith, Minchin, & Belbin, 1987; Legendre & Legendre, 2012).

$$CPC(\mathbf{M}, \hat{\mathbf{M}}) = \frac{2 \sum min(M_{i,j,t}, \hat{M}_{i,j,t})}{\sum M_{i,j,t} + \sum \hat{M}_{i,j,t}} \quad (6)$$

Its distance variant (CPC$_D$) evaluates the accuracy of a predicted migration matrix in reproducing trips at similar distances to the observed data. In this context, *N* represents a histogram, where each bin $N_k$ corresponds to the number of migrants traveling within a specific distance range. Specifically, the range for each bin $k$ is from $2k - 2$ to $2k$ kilometers, effectively categorizing migration distances into 2-kilometer intervals. A score of 0 indicates no migrations at the same distances in the two matrices, while a score of 1 indicates that for any given distance, the predicted and observed migration counts are identical.

$$CPC_D(\mathbf{M}, \hat{\mathbf{M}}) = \frac{2 \sum min(N_k, \hat{N}_k)}{\sum N_k + \sum \hat{N}_k} \quad (7)$$

## RESULTS

[INSERT TABLE 1 HERE]

Table 1 reports average results and 95% confidence bands for each model over several sampled paths of observations. From the table, we observe performance metrics that favor the hierarchical models. According to the metrics, the models are virtually identical, but we conclude that the hierarchical gravity model with varying intercepts (we call this HG$_1$) offers the best performance for its relatively lightweight design. Figure 2 shows credible intervals using HG$_1$ for the out-of-sample prediction of flows for a random sample of 5 state pairs and compares these to point predictions from the homogeneous gravity model.

[INSERT FIGURE 2 HERE]

While the machine learning approaches appear to outperform traditional models, their performance shows considerable variability. This variability may be due, in part, to their observed tendency to assume different model architectures across each of the five paths as a result of cross-validated hyperparameter selection. This illustrates a key advantage of hierarchical approaches over machine learning methods: the more delicate tuning process of machine learning introduces a propensity for human error that is less of a concern with linear methods. Leveraging this advantage, our analysis reveals a crucial insight: when modeling bilateral migration flows, allowing spatial variation in a latent migration parameter can significantly enhance predictive performance.

[INSERT TABLE 2 HERE]

In Table 2, we show that the Bayesian models do not experience the signage errors incurred by the unpooled frequentist approach, demonstrating the ability of the Bayesian hierarchical models to retain the interpretability of (1) and (2).

[INSERT TABLE 3 HERE]

In Table 3, we document the size of each model to demonstrate that the hierarchical model with only a varying intercept achieves excellent predictive performance without excessive computational complexity. Because an overly complex hierarchical model can occasionally thwart the efficacy of Hamiltonian Monte Carlo, in Table 4, we record the extent of degenerate sampling that occurred in the warm-up stage of the NUTS algorithm (Betancourt, 2015). Among the models evaluated, the hierarchical gravity model with varying intercept suffers the least from biased approximation.

[INSERT TABLE 4 HERE]

*Clustering of State Pairs in the Hierarchical Gravity Model*
Since parameters vary over state pairs, the hierarchical models can also inform us about heterogeneities in migrant behavior. In Figure 3, we show that the collection of parameter vectors of $HG_2$ can be distinctly separated into two clusters of state pairs, or contexts. Below the dendrogram are distributions of various context-specific variables, conditioned on the cluster to which the contexts are assigned. In each plot, we record the p-value associated with a $\chi$-test for homogeneity in which the null hypothesis states that the two histograms emerge from the same distribution. In the left plot of the third row in the figure, the orange cluster represents bilateral migration flows of lesser magnitude. In the middle and right plots of the same row, bilateral flows of lesser magnitude are shown to typically correspond to smaller populations and larger distances, respectively. This observation aligns with the relationships suggested by the gravity model. In the bottom row, we report conditional distributions for a few other context-specific variables that separate between each cluster with statistical significance. We observe this phenomenon for the log ratios of the housing index value, land area, and climate-related disaster costs. These histograms clearly illustrate that individuals migrating along a low-flow path are less influenced by the specific characteristics of their destination. Specifically, in the bottom row, the orange histograms representing low-flow paths show greater dispersion compared to the blue histograms for high-flow paths. This suggests that individuals migrating along low-flow paths may migrate for more nuanced reasons than those on high-flow paths, who likely respond more directly to destination characteristics.

In addition to these three variables, we ran the homogeneity test for several other context-specific variables, including the log ratio of median household income, population density (measured by population divided by land area), and affordability (measured by median household income divided by housing index value). These other variables, however, gave no evidence suggesting the two histograms were from different distributions.

[INSERT FIGURE 3 HERE]

One key insight from Figure 3 is that variables whose conditional histograms show considerable variation across different conditions or subgroups are likely important to include as predictor variables in a new hierarchical model that allows their effects to vary spatially or across different contexts. Furthermore, we estimated "generalized" hierarchical gravity and radiation models. These are similar to the models discussed in this paper, but instead of using $P_i$ and $P_j$ directly, they incorporate a linear combination of several socioeconomic features (Alis, Legara, & Monterola, 2021; Kim & Cohen, 2010). Hence, this class of models offers flexibility by allowing the inclusion of many exogenous variables in addition to population and distance–such as median household income, population density, housing costs, affordability, land area, and disaster costs. While this seemed a promising direction for addressing the effects of variables beyond just population and distance, the resulting models turned out to be overspecified, due to the prevalence of overfitting and unstable parameter estimation. As a result, such generalized models were not pursued further.

*Technical Challenges in Hierarchical Modeling*

While the hierarchical framework can be powerful for modeling heterogeneities, ensuring proper approximate inference is not a trivial task. In a Gaussian hierarchical model with such high dimensionality, a lack of information in each context can hinder the identifiability of posterior densities. Specifically, a poorly informed scale parameter can result in a highly curved likelihood that can cause divergent trajectories in Hamiltonian Monte Carlo and induce biased sampling (Betancourt, 2020). To mitigate the lack of information in data-sparse contexts, upsampling the available data within those contexts emerges as a natural strategy. We approximated hierarchical models using training data that was resampled five times, giving each site at most 60 observations to learn from. Under the same estimation procedure as for the smaller dataset, we observed that for each model, the proportions of warm-up samples that were divergent were the same as in Table 4.

The failure of off-the-shelf reparameterization techniques and upsampling to corral these degeneracies suggests limitations of the data size or the model specification outlined in Figure 1. By taking the traditional models hierarchical in the way that we have, we assume that all context-specific variables are conditionally independent given the common coefficient. However, it could very well be the case that there are dependencies in the second tier of the hierarchical model that are not accounted for by our current specification, which warrants either an introduction of additional latent variables or a full rethinking of the proposed Bayesian network. Welch and Raftery (2022), for example, introduce a hierarchical model for migration flows that is semi-parametric and consists of multiple hierarchical tiers, which may help alleviate degeneracies brought on by a Gaussian specification and a violation of conditional independence assumptions. However, as their model is endogenous and does not rely on exogenous features, it cannot provide the impetus for identifying the driving factors for migration. Hence, a direction worthy of further study may be to reformulate the Bayesian networks used in this paper so that they better accommodate unbiased sampling as well as a large collection of variables.

## CONCLUSIONS

Predicting both the origin and destination of individuals 'movements poses a fundamental challenge within the field of human mobility research. The hierarchical Bayesian model framework that we propose outperforms recent machine learning approaches for this task while maintaining the interpretability of traditional methods. In essence, by accounting for spatial variation in the drivers of migration decisions, the model is better equipped to capture the dynamics of bilateral migration flows. In particular, merely permitting spatial variation in a parameter governing latent migration flows results in a substantial increase in predictive $R^2$. We show this through a case study on migrant flows in the United States, and we posit that this framework can be readily adapted for analyses on any geospatial scale. In addition to enhancing predictive performance, the hierarchical models reveal insights about heterogeneities in migrant preferences. Our clustering analysis of spatially varying parameters reveals evidence suggesting that individuals migrating along low-flow paths are more nuanced in their decision-making than individuals migrating along high-flow paths. While the hierarchical models demonstrate success in predictive performance, we also document difficulties with the approach. Specifically, current Bayesian estimation procedures face difficulties in mitigating biased approximations that can arise when scaling to excessively large models involving latent variables that potentially violate conditional independence assumptions.

Our results lend further support to the use of machine learning for migration modeling, as the neural network and decision tree models we evaluate demonstrate superior predictive performance compared to the traditional linear methods. As high-resolution socioeconomic data becomes increasingly accessible in countries that also monitor human migration flows, there is a growing opportunity to develop tailored machine-learning models that can deepen our understanding of human migration dynamics. While machine learning models hold promise for predicting migration flows, their development should proceed with caution due to the lack of intuitive guidance for selecting an appropriate architecture. However, leveraging Bayesian statistics can provide uncertainty quantification around latent variables, enabling more principled inferences to be made. For further research directions, we propose exploring a redesigned Bayesian network that could more effectively accommodate a large number of variables while mitigating biased approximations, thereby enabling more accurate quantification of heterogeneities and regression parameters.

## ACKNOWLEDGMENTS

The research reported here was supported by the Army Research Office/Army Research Laboratory under award W911NF1810267 (Multidisciplinary University Research Initiative). The funders had no role in study design, data collection and analysis, decision to publish, or preparation of the manuscript. The views and conclusions contained in this document are those of the authors and should not be interpreted as representing the official policies either expressed or implied of the Army Research Office or the U.S. Government.

## CONFLICT OF INTEREST STATEMENT

The authors declare no conflict of interest.

## DATA AVAILABILITY STATEMENT

The data that support the findings of this study are publicly available from the United States Census Bureau (https://www.census.gov), Federal Housing Finance Agency (https://www.fhfa.gov), National Centers for Environmental Information (https://www.ncei.noaa.gov), and Federal Reserve Bank of St. Louis (https://fred.stlouisfed.org). All code used for data processing, analysis, and visualization will be made available on GitHub (URL to be provided upon publication). The code includes implementation of the hierarchical Bayesian models, machine learning comparisons, and clustering analyses described in this study.

# Tables

**Table 1.** Comparison of out-of-sample model performance. We use the acronyms HG and HR to denote hierarchical gravity and hierarchical radiation results, respectively, and we subscript each acronym with 1 and 2 to denote whether the intercept is varying or both the intercept and coefficients are varying, respectively. In bold are the best values per metric.

| MODEL | MAE | $R_2$ | CPC | $CPC_D$ |
|---|---|---|---|---|
| Gravity | 1,894 ± 7 | 0.402 ± 0.001 | 0.634 ± 0.001 | 0.756 ± 0.001 |
| Radiation | 1,658 ± 6 | 0.527 ± 0.003 | 0.695 ± 0.001 | 0.817 ± 0.002 |
| XGBoost | 1,306 ± 23 | 0.767 ± 0.016 | 0.785 ± 0.003 | 0.943 ± 0.002 |
| ANN | 1,436 ± 44 | 0.677 ± 0.032 | 0.748 ± 0.014 | 0.873 ± 0.033 |
| $HG_1$ (intercept varying) | **1,077** ± 14 | 0.828 ± 0.006 | **0.827** ± 0.002 | 0.964 ± 0.002 |
| $HG_2$ (intercept + coef varying) | 1,080 ± 14 | **0.829** ± 0.006 | **0.827** ± 0.002 | **0.967** ± 0.002 |
| $HR_1$ (intercept varying) | 1,079 ± 14 | 0.827 ± 0.006 | 0.825 ± 0.002 | 0.959 ± 0.002 |
| $HR_2$ (intercept + coef varying) | 1,079 ± 14 | 0.828 ± 0.006 | 0.826 ± 0.002 | 0.961 ± 0.002 |

**Table 2.** Percent of site-specific effects whose 90% credible intervals are in sign agreement with the original model. To demonstrate the efficacy of the Bayesian framework's shrinkage property, we compare the signage of coefficients between Bayesian hierarchical models and unpooled models estimated using ordinary least squares. We omit the intercept term ($\alpha$) from the table because, while the original gravity (1) and radiation (2) models imply it should be negative, the concept of latent migration in our hierarchical models doesn't inherently have a positive or negative interpretation. The coefficients $\beta_{1:i,j}$ (origin population), $\beta_{2:i,j}$ (destination population), $\beta_{3:i,j}$ (distance), and $\beta_{4:i,j}$ (intervening opportunities) should align in sign with the intuition of the gravity (1) and radiation (2) models. Note that $\beta_{4:i,j}$ is only applicable to radiation models as it represents intervening opportunities, which are not part of gravity model formulation. This alignment demonstrates the ability of Bayesian hierarchical models to retain the interpretability of the original models while accounting for spatial heterogeneity.

| Model | $\beta_{1:i,j}$ (Origin Pop.) | $\beta_{2:i,j}$ (Dest. Pop.) | $\beta_{3:i,j}$ (Distance) | $\beta_{4:i,j}$ (Interv. Opp.) |
|---|---|---|---|---|
| $HG_2$ | 100% | 100% | 100% | – |
| $HR_2$ | 100% | 100% | 100% | 100% |
| Unpooled gravity | 49.80% | 47.92% | 46.08% | – |
| Unpooled radiation | 51.37% | 46.24% | 52.82% | 47.45% |

**Table 3.** We refer to model size as the number of trainable parameters in a model. For the ML models, cross-validated grid search in each of the 5 paths leads to models of different sizes, so we report the average model size. We exclude hyperparameters such as step sizes in HMC and learning rates in the ML optimization algorithms. For the Bayesian models, we only count parameters that appear in the posterior predictive mean, which are the intercepts and coefficients.

| Model | Model Size |
|---|---|
| Gravity | 4 |
| Radiation | 5 |
| XGBoost | 2,634 |
| ANN | 1,554 |
| $HG_1$ | 2,553 |
| $HG_2$ | 10,200 |
| $HR_1$ | 2,554 |
| $HR_2$ | 12,750 |

**Table 4.** Efficacy and efficiency of estimation procedures. Top: Percentage of 250 warm-up samples resulting in divergent transitions. Bottom: Total inference time (in minutes) for 250 warm-up samples and 1000 posterior samples. VIP refers to our variational inference preprocessing approach. Results show that VIP generally reduces divergent transitions and speeds up inference.

| Model | VIP *(minutes)* | No VIP *(minutes)* | Model | VIP *(minutes)* | No VIP *(minutes)* |
|---|---|---|---|---|---|
| $HG_1$ | **1.6%** | 2.4% | $HG_1$ | **6.4** | 38.1 |
| $HG_2$ | **1.6%** | 1.6% | $HG_2$ | 11.8 | 43.6 |
| $HR_1$ | 2.0% | 2.0% | $HR_1$ | 49.6 | 69.2 |
| $HR_2$ | **1.6%** | 2.0% | $HR_2$ | 36.2 | 62.0 |

# Figures

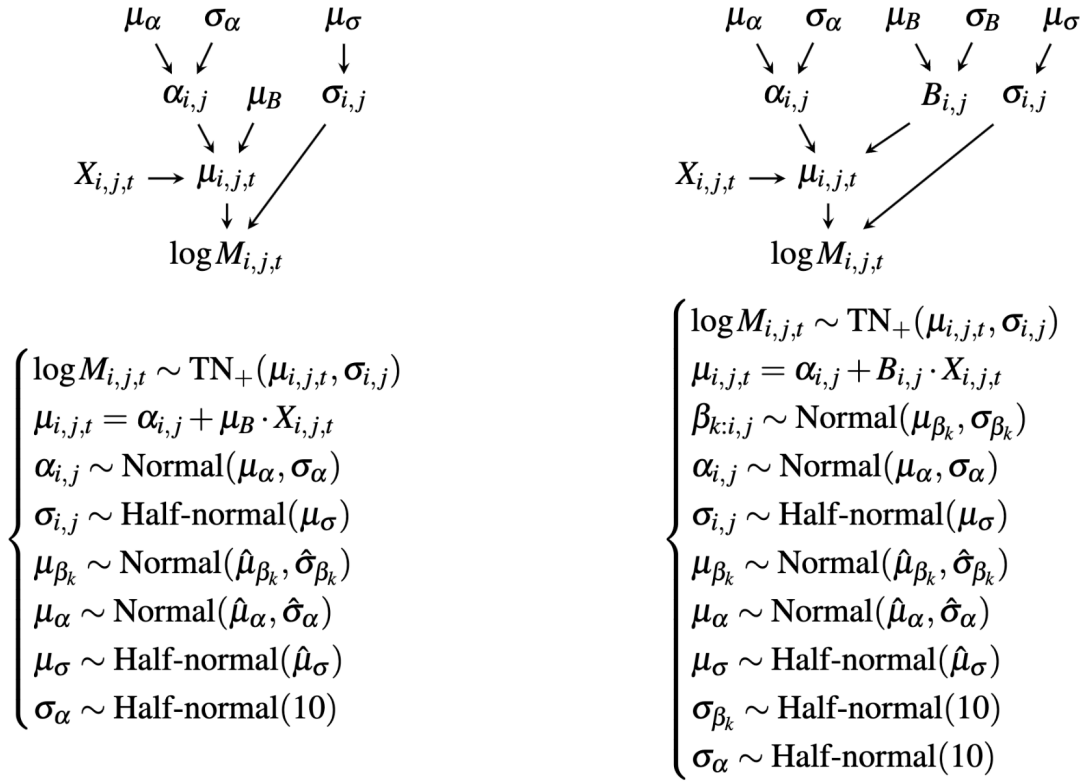

(A) Hierarchical model w/ varying intercept     (B) Hierarchical model w/ all parameters varying

**Figure 1.** Graphical representation of competing hierarchical models. For the gravity and radiation models, $X_{i,j,t} = \log P_{i,t}, \log P_{j,t}, \log D_{i,j}$ and $X_{i,j,t} = \log P_{i,t}, \log P_{j,t}, \log(P_{i,t} + S_{i,j,t}), \log(P_{i,t} + P_{j,t} + S_{i,j,t})$, respectively. $\mathrm{TN}+$ denotes a truncated normal distribution with a support of non-negative real numbers. The model includes several terms: $\mu_{i,j,t}$ is the expected migration flow from location $i$ to location $j$ at time $t$; $\sigma_{i,j}$ denotes the standard deviation of the migration flow between locations $i$ and $j$; $\alpha_{i,j}$ is the intercept term specific to the pair of locations $i$ and $j$; $\mu_B$ is the mean of the slope parameters for the regression on features $X_{i,j,t}$; $\mu_\alpha$ represents the mean of the intercepts $\alpha_{i,j}$; $\sigma_\alpha$ is the standard deviation of the intercepts $\alpha_{i,j}$; $\mu_\sigma$ is the mean of the standard deviations $\sigma_{i,j}$; $\beta_{k,i,j}$ is the slope parameter for the $k$-th feature specific to the pair of locations $i$ and $j$; $\sigma_{\beta_k}$ denotes the standard deviation of the slope parameters $\beta_{k,i,j}$; and $X_{i,j,t}$ is the feature set including variables $\log P_{i,t}$, $\log P_{j,t}$, and $\log D_{i,j}$. $\sigma$ represents the standard deviation, capturing the variability or spread of the distribution of a parameter. $\mu_\sigma$ is the mean of the standard deviation $\sigma_{i,j}$, characterizing the central tendency of the standard deviations across different contexts. Arrows in the diagram indicate the hierarchical relationships: for example, $\mu_\alpha$ and $\sigma_\alpha$ influence $\alpha_{i,j}$, $\mu_B$ influences $\beta_{k,i,j}$, and $\mu_\sigma$ influences $\sigma_{i,j}$.

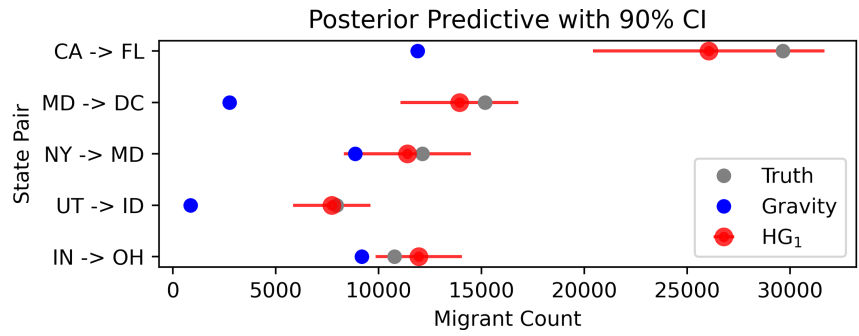

**Figure 2.** Out-of-sample predictions for 5 randomly sampled state pairs. We compare 90% credible intervals drawn from the posterior predictive distribution for $HG_1$ to truth values and to point predictions from the gravity model. Predictions are averaged across the three out-of-sample years 2017-2019.

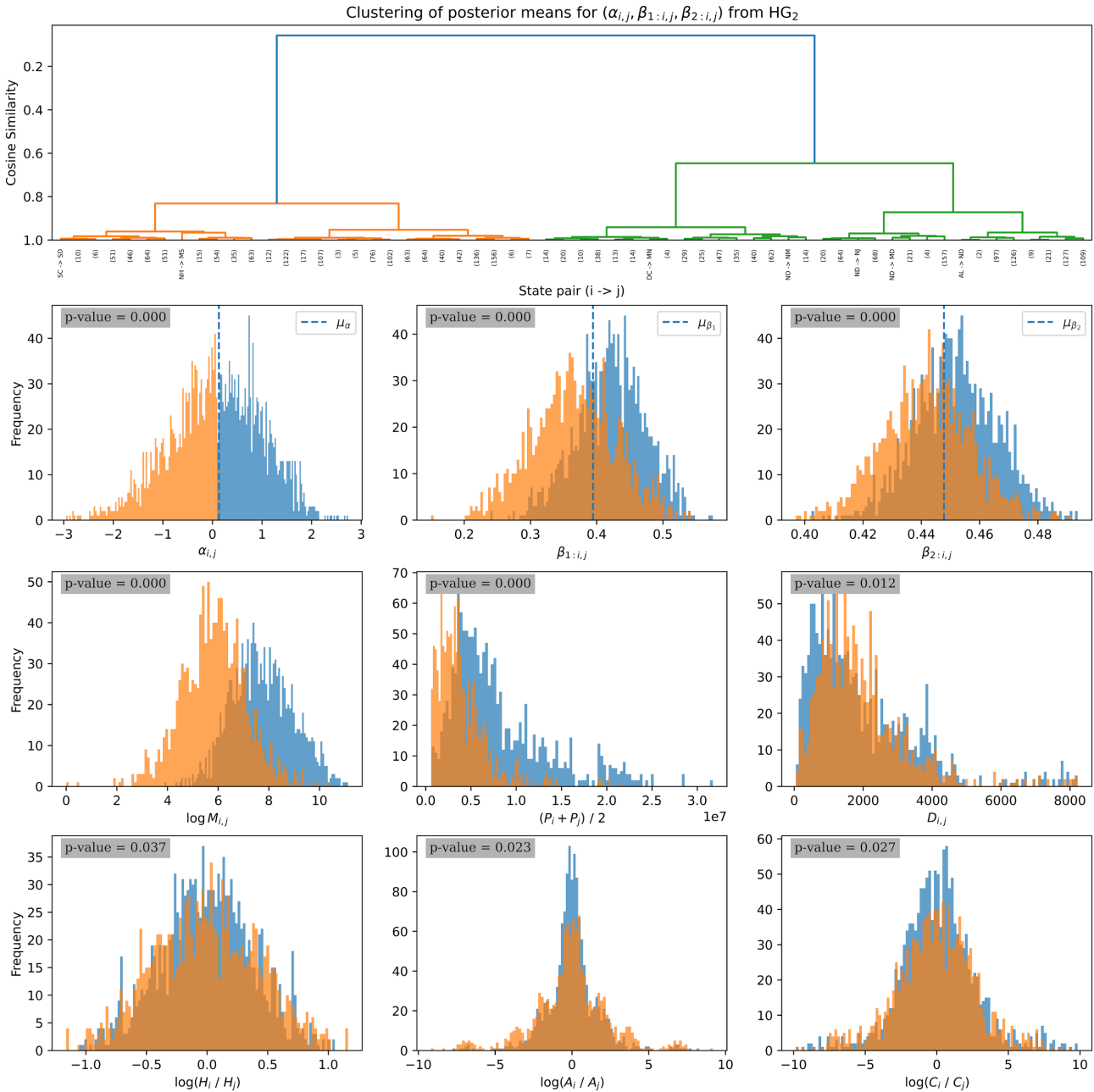

**Figure 3.** Clustering of state pairs according to posterior means of parameter vectors for HG$_2$. The figure in the first row is a dendrogram clustering the collection of vectors $\{\alpha_{i,j}, \beta_{1:i,j}, \beta_{2:i,j}\}$ according to cosine similarity. The second row demonstrates how the two main clusters separate among the parameters. The left plot of the third row shows that the orange cluster corresponds to low-flow migration paths. The bottom row shows the conditional distributions for the log ratio of housing index value ($\log(H_i/H_j)$), land area ($\log(A_i/A_j)$), and climate-related disaster costs ($\log(C_i/C_j)$). In each plot is a p-value obtained from a $\chi^2$-test for homogeneity, where the null hypothesis is that the two histograms emerge from the same distribution. The omission of $t$ in the subscript of time-varying variables indicates the variable is averaged over the 15 years of data.